\documentclass[aps,prl,twocolumn,showkeys,groupedaddress,nofootinbib]{revtex4}
\usepackage[pdftex]{graphicx}

\begin{document}
\title{A flux tunable superconducting quantum circuit based on Weyl semimetal MoTe$_{2}$}

\author{K. L. Chiu$^{\dagger,*}$, D. G. Qian$^{\ddagger}$, J. W. Qiu$^{\ddagger}$, W. Y. Liu$^{\ddagger}$, D. Tan$^{\ddagger}$, V. Mosallanejad$^{\mp}$, S. Liu$^{\ddagger}$, Z. T. Zhang$^{\ddagger}$, Y. Zhao$^{\ddagger}$, D. P. Yu$^\ddagger$}

\affiliation{$^\dagger$Department of Physics, National Sun Yat-Sen University, Kaohsiung 80424, Taiwan}
\affiliation{$^\ddagger$Shenzhen Institute for Quantum Science and Engineering, Southern University of Science and Technology, Shenzhen 518055, China}
\affiliation{$^\mp$Key Lab of Quantum Information, University of Science and Technology of China, Hefei 230026, China}

\date{\today}

\begin{abstract}
Weyl semimetals for their exotic topological properties have drawn considerable attention in many research fields. When in combination with s-wave superconductors, the supercurrent can be carried by their topological surface channels, forming junctions mimic the behavior of Majorana bound states. Here, we present a transmon-like superconducting quantum intereference device (SQUID) consists of lateral junctions made of Weyl semimetal Td-MoTe$_2$ and superconducting leads niobium nitride (NbN). The SQUID is coupled to a readout cavity made of molybdenum rhenium (MoRe), whose response at high power reveal the existence of the constituting Josephson junctions (JJs). The loop geometry of the circuit allows the resonant frequency of the readout cavity to be tuned by the magnetic flux. We demonstrate a JJ made of MoTe$_2$ and a flux-tunable transmon-like circuit based on Weyl materials. Our study provides a platform to utilize topological materials in SQUID-based quantum circuits for potential applications in quantum information processing. 
\end{abstract}


\maketitle
\def\thefootnote{*}\footnotetext{Corresponding author: K. L. Chiu (klc@mail.nsysu.edu.tw)}\def\thefootnote{\arabic{footnote}}
\section{I. Introduction}

Weyl semimetals are three-dimensional phases of matter with gapless band structures that are protected by topology and symmetry \cite{Armitage2018}. The low-energy bands in Weyl semimetals can be viewed as a three-dimensional version of graphene, i.e., dispersing linearly along all the three momentum directions across the Weyl points (WPs). The WPs always appear in pairs with opposite chirality, and are connected by open curve-like Fermi surfaces with nondegenerate spin texture, as known as surface Fermi arcs \cite{Armitage2018}. Recently, orthorhombic T$_d$-phase transition metal dichalcogenides (such as WTe$_2$ and MoTe$_2$) have been predicted to be type-II Weyl semimetals \cite{Soluyanov2015,Chang2016}, which is later supported by several experimental studies \cite{Jiang2017,Li2017}. In addition, the topologically protected surface states of Weyl semimetals are expected to transmit electrical currents without backscattering, and therefore could provide a robust weak link to carry supercurrent in Josephson junctions (JJ). A recent study on a lateral junction made of T$_d$-WTe$_2$ has shown a nontrivial temperature dependence of critical current $I_c$($T$), in which a short junction behavior was observed in a long junction device \cite{Shvetsov2018}. The long-lived temperature dependence of $I_c$($T$) has been attributed to the topologically protected surface Fermi arc states in Weyl semimetal \cite{Shvetsov2018}. In another study, a missing $n$=1 $Shapiro$ step, indicating the existence of the nontrivial 4$\pi$-periodic supercurrent, was observed in a nanowire JJ made of Dirac semimetal Cd$_3$As$_2$ and is also attributed to its surface Fermi arc states \cite{Wang2018a}. On the other hand, integrating 2D materials (such as graphene) with superconducting circuits is an emerging topic in searching of new types of quantum computing devices owing to its superb conductivity and 2D gateable nature \cite{Casparis2018,Chiu2018,Kroll2018a,Wang2019}. Several key observations, such as gate-tunable qubit energy \cite{Kroll2018a,Wang2019}, Rabi oscillation and qubit relaxation time $T_1$ (dephasing time $T^\ast_2$) at the scale of 36 ns (51 ns), have been reported \cite{Wang2019}. 

Topological materials, for their topologically protected surface and edge states which can serve as a robust channel to carry supercurrent, are also promising candidates for use in 2D materials-based quantum computing devices \cite{Chiu2017,Chien2020}. Recently, a topological insulator nanoribbon has been reported for use in superconducting quantum circuits \cite{Schmitt2020}. In addition, the S-T-S junction (S is superconductor and T is topological material) naturally provides a platform to explore the physics associated with Majorana bound states (MBS) \cite{Fu2009,Fu2008,Badiane2013}. In this letter, we explore a flux-tunable superconducting quantum intereference device (SQUID) consists of Weyl semimetal Td-MoTe$_2$ and superconducting leads niobium nitride (NbN). The S-T-S junction is capacitively coupled to a superconducting readout cavity made of molybdenum rhenium (MoRe), with an intended design similar to a transmon. The power dependence of the readout cavity frequency demonstrates the successfully fabricated JJs while the flux-tunability of that reveal a symmetric DC SQUID. We compare our results with a theoretical framework where 4$\pi$-periodic supercurrent plays a role in a topological junction. Our results demonstrate a flux-tunable DC SQUID based on Weyl semimetal and the integration with readout cavity provides a platform for potential applications in quantum information processing.

\begin{figure}[!t]	
\includegraphics[scale=0.41]{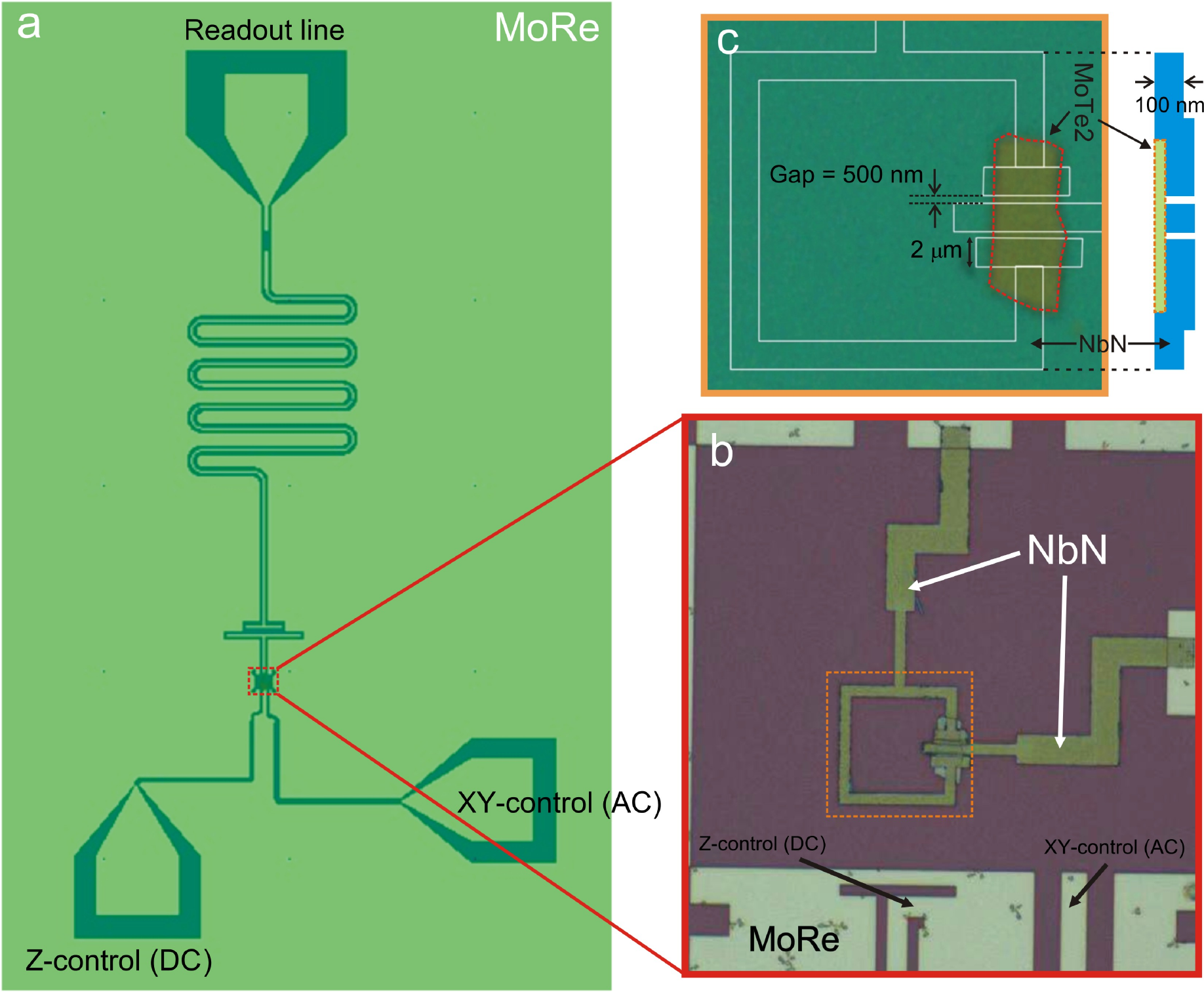}
\caption{(a) Optical micrograph of the MoTe$_2$ transmon-like SQUID device. The MoTe$_2$ SQUID is shunted by the T-shaped island to the surrounding ground plane and coupled to a reflective readout cavity made of MoRe. The SQUID and shunting capacitor is equipped with a DC line (Z-control) for flux tuning and a AC line (XY-control) for applying additional drive tone. (b) Optical micrograph showing the MoTe$_2$ SQUID made of NbN, and the Z-control and XY-control line made of MoRe. (c) The geometry of MoTe$_2$ SQUID before the deposition of NbN. The SQUID is formed by a square loop (line width: 2 $\mu$m) with MoTe$_2$ covering two 500 nm gaps. The right inset shows the cross-section across the gaps.}      
\label{Fig1}
\end{figure}

\section{II. Device fabrication and measurement setup}
Fig. \ref{Fig1} shows the geometry of the SQUID along with the surrounding superconducting quantum circuits in our device. The SQUID is made of two NbN-MoTe$_2$-NbN junctions, with a enclosed loop defined by a square (18 $\times$ 18 $\mu m^2$) of NbN. The SQUID is shunted by a capacitor formed between a T-shape island and the surrounding ground, which are made by MoRe. The SQUID and shunting capacitor is equipped with three other components: a reflective microwave readout cavity, a DC line for current-induced flux control and a AC line for applying additional drive tones, forming a transmon-like device as shown in Fig. \ref{Fig1}(a). The device fabrication consists of three steps. Firstly, 80 nm MoRe is deposited on an undoped SiO$_2$/Si substrate, followed by a photolithography (laser writer) and Inductively Coupled Plasma Etching (ICP dry etching) to define the quantum circuit as shown in Fig. \ref{Fig1}(a). Secondly, the fabrication of van der Waals layered materials is based on the polymer-free dry transfer method \cite{R.2010,Mayorov2011,Haigh2012,Wang2013}, in which a poly-carbonate (PC) is used as a stamp to pick up the exfoliated MoTe$_2$ (bulk purchased from HQgraphene), followed by a transfer onto the desired area (the red dashed square in Fig. \ref{Fig1}(a)) for further processing. Thirdly, a Electron-beam lithography and sputtering of 100 nm NbN are used to fabricate the SQUID and the leads connecting the T-shape island and ground, as shown in Fig. \ref{Fig1}(b). Fig. \ref{Fig1}(c) shows the zoom-in image of the JJs before the sputtering of NbN, in which the gap of 500 nm is the designed junction length. More details of the fabrication process can be found in section I in the supplementary material. 

The measurement scheme exploited in this work is illustrated in Fig. \ref{FigS2}. The sample is mounted in a aluminum box and thermally anchored to a dilution refrigerator with a base temperature of 10 mK. The readout cavity of device is connected to the input and output of a network analyzer for scattering parameters (S$_{21}$) measurement. A Digital-to-Analog Converter (DAC) used for applying DC voltage is connected to the Z-control line for magnetic flux tuning, while a RF source used for applying additional tone for two-tone measurements is connected to the XY-control line [Fig. \ref{FigS2}]. More details of the measurement scheme and the associated components can be found in section II in the supplementary material.

\begin{figure}[!t]	
\includegraphics[scale=0.43]{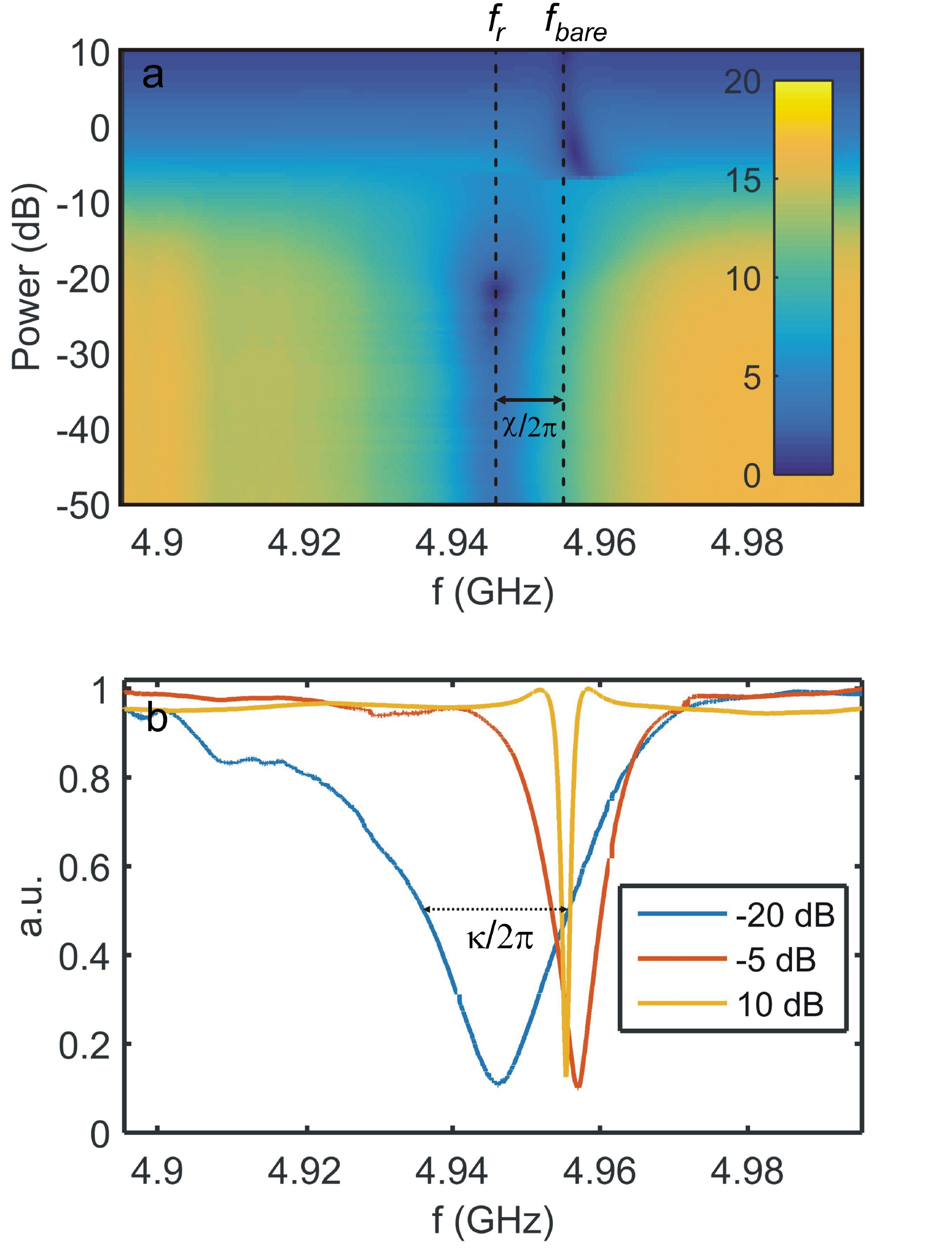}
\caption{(a) The power spectrum measured by recording $S_{21}$ as a function of power and frequency applied to the readout-in line. The existence of JJs in the SQUID is characterized by a cavity dispersive shift $\chi/2\pi$. The spectrum peaks at 4.945 GHz at low power, then undergoes a broad regime before the peak re-appear at P $\approx$ -5 dB, and finally enters a bright regime at P = 10 dB. (b) Linecuts in (a) at three different powers (data normalized to 1). The full width at half maximum of the peak indicates the cavity decay rate $\kappa/2\pi$.}
\label{Fig2}
\end{figure}

\section{III. Device characterization and results}
The SQUID along with the shunting capacitor is coupled to a $\lambda$/2 superconducting reflective cavity with a bare resonant frequency $f_c$ $\approx$ 4.955 GHz and a loaded quality factor Q $\approx$ 3964 obtained $via$ $f_c$/$\Delta f$ from the trace of P = 10 dB in Fig. \ref{Fig2}(b). We first performed a standard qubit power-dependent experiment, in which the S$_{21}$ is measured as a function of the input power applied to the readout line, as shown in Fig. \ref{Fig2}(a). This is a conventional way to confirm the existence of JJ in transmon\cite{Reed2010}, where the resonant frequency of the cavity shifts toward the transmon frequency at large enough power (here at around -5 dB). Here, we note that we have not demonstrated temporal coherence or energy-level spectroscopy in our device, which does not justify the device as a transmon. However, for simplicity we still use terms, such as transmon frequency or transmon-cavity coupling, in the following content. The power-dependent frequency shift is generally attributed to the nonlinearity that cavity inherits from JJs, in which cavity can not be approximated to a perfect harmonic oscillator as the photon occupation number increases \cite{Reed2010}. Fig. \ref{Fig2}(b) shows the linecuts at different powers. At lower power (-20 dB), the cavity frequency centers at 4.945 GHz with an asymmetric line shape, indicating a canonical behavior of a Kerr-Duffing oscillator \cite{Reed2014a}. The peak exists until P $\approx$ -10 dB, beyond which the cavity response undergoes a broad regime (see section III in the supplementary material). Beyond a critical power (P $\approx$ -5 dB), the cavity response re-appears, with a narrower line shape of Lorentzian and a frequency shifted to 4.957 GHz, as shown in Fig. \ref{Fig2}(b). As the power further increases, the cavity linewidth continue to decrease, which is expected for a MoRe cavity in an elevated power \cite{Singh2020}. However, while the power is increased, surprisingly the peak position still moves slightly with power, eventually reaching a stable value of around 4.955 GHz at P = 10 dB. This is known as the bare cavity frequency at which the cavity would resonate in the absence of transmon’s JJs. This effect is more pronounced in Fig. \ref{FigS3}, where we have reduced the attenuation on the readout-in line to apply larger powers. The reason for the post-bright-regime frequency shift is not fully decisive. We suspect it could be due to the residual nonlinearity (in the range of -5 dB $\leq$ P $\leq$ 10 dB) that cavity still inherited from one junction, while the other one was saturated earlier at a relatively lower power.

\begin{figure*}[!t]	
\includegraphics[scale=0.71]{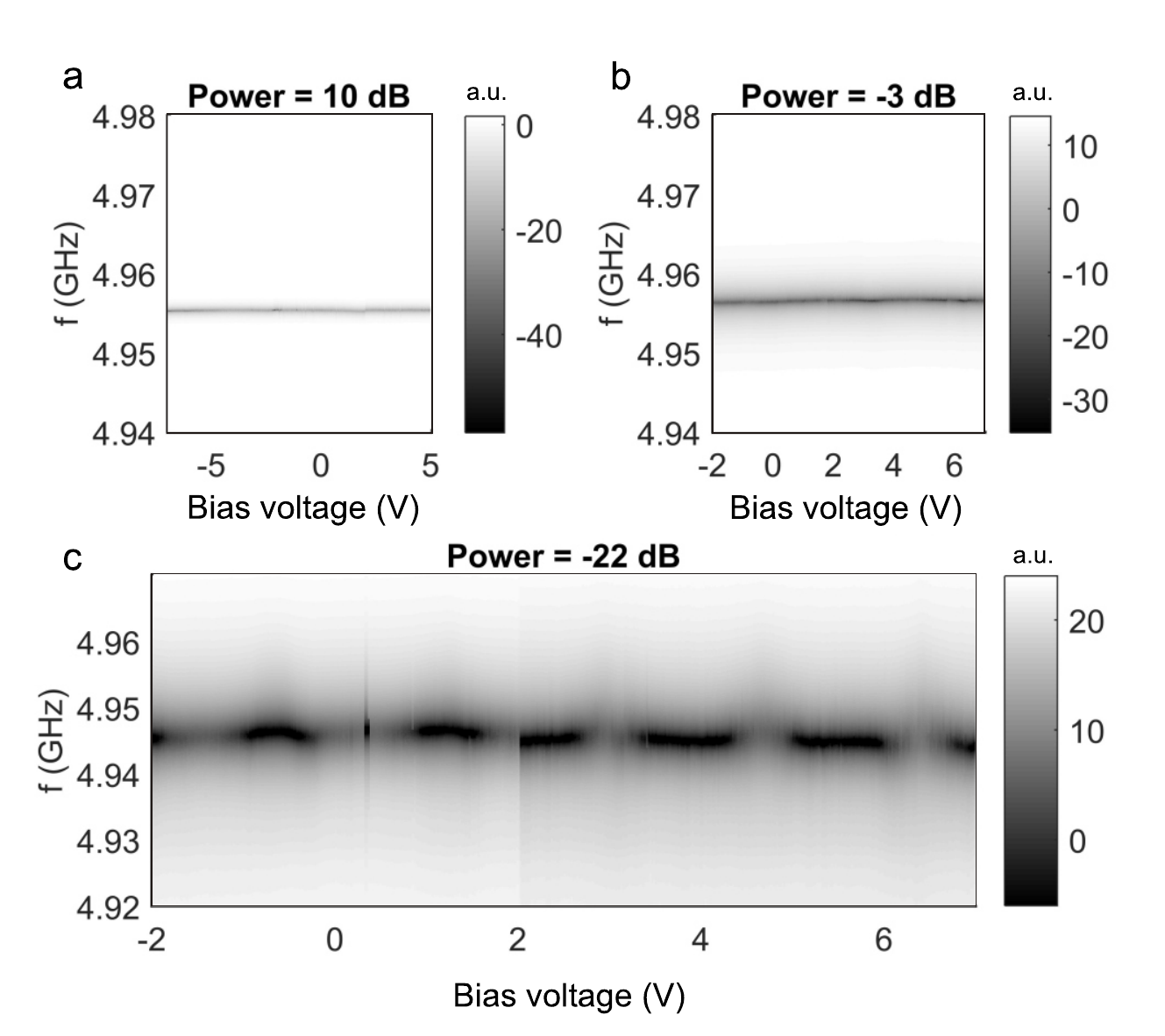}
\caption{Cavity spectrum measured by recording $S_{21}$ as a function of readout-in frequency and bias voltage (magnetic field) with readout-in power at (a) P = 10 dB (b) P = -3 dB (c) P = -22 dB. The flux modulation is absent at higher powers where the cavity has passed the critical point in power spectrum.}
\label{Fig3}
\end{figure*}

The magnitude of the shift $\chi/2\pi$ between cavity's frequency at low power ($f_r$) and at high-power ($f_{bare}$), as shown in Fig. \ref{Fig2}(a), allows us to estimate the transmon frequency via $\chi = g^2/\Delta$, where $g$ is the transmon-cavity coupling strength and $\Delta/2\pi = f_r - f_t$ is the difference between $f_r$ and the transmon frequency $f_t$. In our system, $g/2\pi$ is estimated to be 116 MHz (see section IV in the supplementary material), from which we can infer $f_t$ $\approx$ 6.29 GHz. The fact that $\left|g/\Delta\right|$ $\ll$ 1 indicates our transmon is working in the dispersive regime. However, $\left|g^2/\Delta\right|$ is close to the cavity decay rate $\kappa$ [see Fig. \ref{Fig2}(b)], which does not satisfy the strong dispersive condition of $\left| \frac{g^2}{\Delta} \right|$ $\gg$ $\kappa$ and may in fact place a limitation for us to perform the state-dependent readout of cavity. The relaxation time T$_1$ set by Purcell effect can be expressed as $T_1 = \frac{1}{\gamma_{pur}} = \frac{\Delta^2}{\kappa g^2}$, where $\gamma_{pur}$ is the rate at which the transmon decays due to the Purcell effect \cite{Koch2007,Houck2008}. Substituting $\kappa/2\pi$ $\approx$ 20 MHz, $g/2\pi$ $\approx$ 116 MHz and $\Delta/2\pi$ = $\left|f_r - f_t\right|$ = 6.29 $-$ 4.945 = 1.345 GHz, we estimate an upper bound of $T_1$ $\approx$ 1 $\mu$s. In an attempt to measure the relaxation time, we have performed the two-tone measurement, in which an additional drive tone searching for transmon frequency is applied to the XY-control line (see section V in the supplementary materials). However, no state-dependent cavity response was observed, indicating a very short coherence time. In future endeavor, we anticipate an improved Q-factor in readout cavity could suppress the Purcell decay and thus enhance $T_1$. In addition, by replacing SiO$_2$ dielectric layer, careful transfer of 2D materials to form cleaner JJ interfaces, and increased magnetic and infrared radiation shielding, $T_1$ and $T_2^*$ are expected to be further improved \cite{Oliver2013,OConnell2008,Corcoles2011,Barends2011,Place2020}. Transmon frequency $f_t$ is the transition frequency from the ground state to the first excited state of the transmon and can be expressed as $f_t$ = $E_{01} /h$ $\approx$ $\sqrt{8 E_J E_C}/h$, where $E_C$ = $\frac{e^2}{2C_\Sigma}$ is the charging energy and $E_J$ = $\frac{\Phi_0I_C}{2\pi}$ is the Josephson energy ($\Phi_0$ = $h/2e$ is the flux quantum). From electrostatic simulations we estimate a charging energy of $E_C/h$ $\approx$ 222 MHz ($C_\Sigma$ $\approx$ 87 fF, see section IV in the supplementary material). Combined with $E_{01} /h$ = 6.29 GHz, we can infer the Josephson energy $E_J/h$ $\approx$ 22 GHz and the critical current of the SQUID $I_C$ $\approx$ 44.2 nA. By changing the magnetic flux threading the loop of SQUID, we are able to change the critical current, hence $E_J$ and $f_t$, which leads to different $\chi$ to be measured. The flux-tunability of cavity frequency allows us to distinguish the SQUID operation regime from the cavity bright state. Fig. \ref{Fig3} shows the resonant frequency of cavity as a function of flux bias at different readout-in powers. At lower power (P = -22 dB), cavity peak shows a periodic modulation with magnetic flux, while at higher powers (P = -3 dB and 10 dB), cavity frequency presents no flux modulation as JJs acting as absent due to entering the bright regime.

\begin{figure*}[!t]	
\includegraphics[scale=0.75]{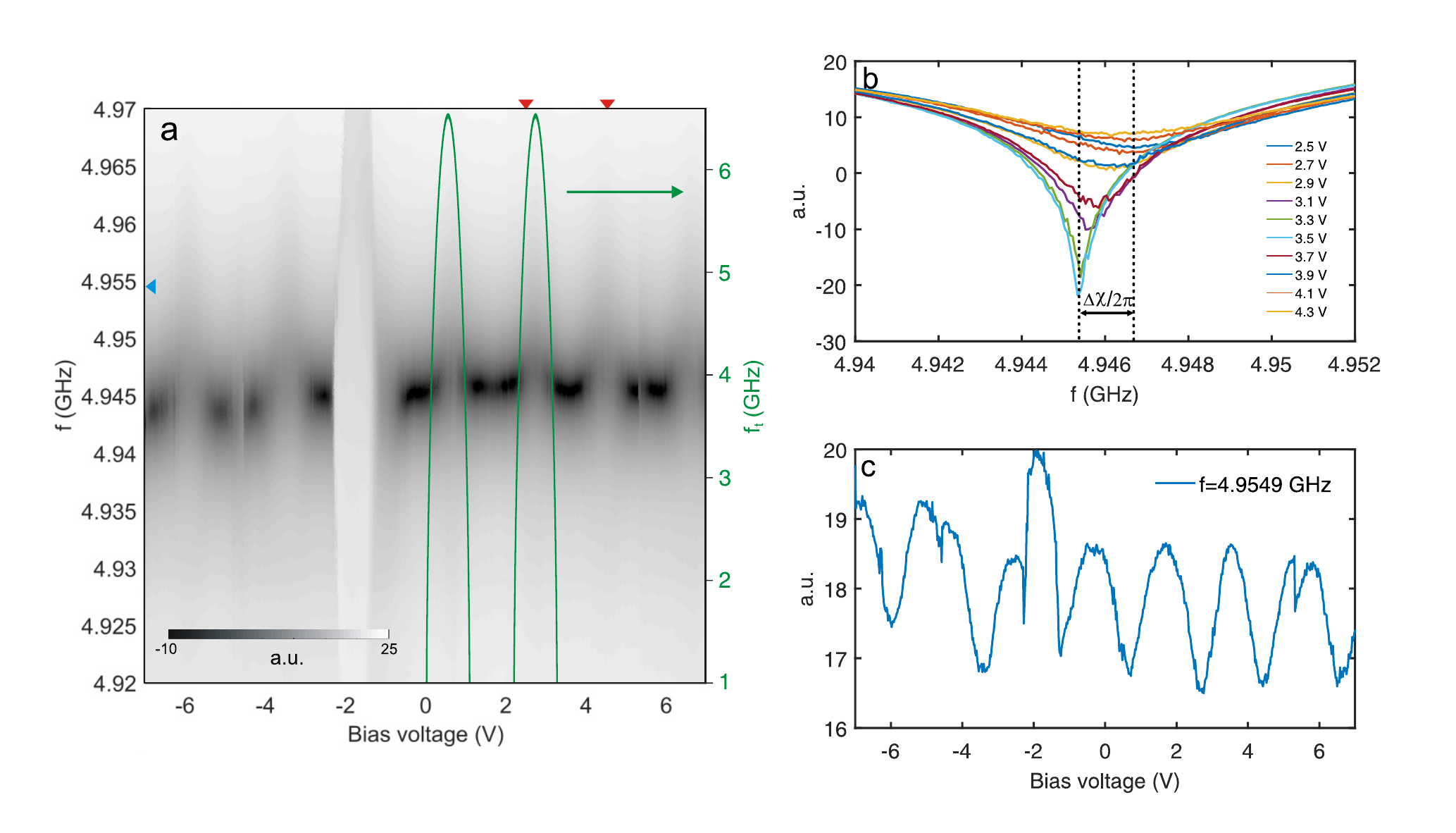}
\caption{(a) $S_{21}$ as a function of readout-in frequency and bias voltage at P = -20 dB showing the periodic flux tuning. The green solid lines denote the inferred transmon transition frequency $f_t$ as depicted in the main text. (b) Linecut along the frequency direction from V = 2.5 V to V = 4.3 V [range marked by red triangles in (a)]. (c) Linecut along the bias voltage direction at $f$ = 4.9549 GHz [marked by blue triangle in (a)]. Note that the bias voltage range from -7 V to 7 V corresponds to a current range from -4.66 mA to 4.66 mA, as a 1.5 k$\Omega$ resistor has been use to generate a DC current for flux tuning.}
\label{Fig4}
\end{figure*}

\begin{figure*}[!t]	
\includegraphics[scale=0.75]{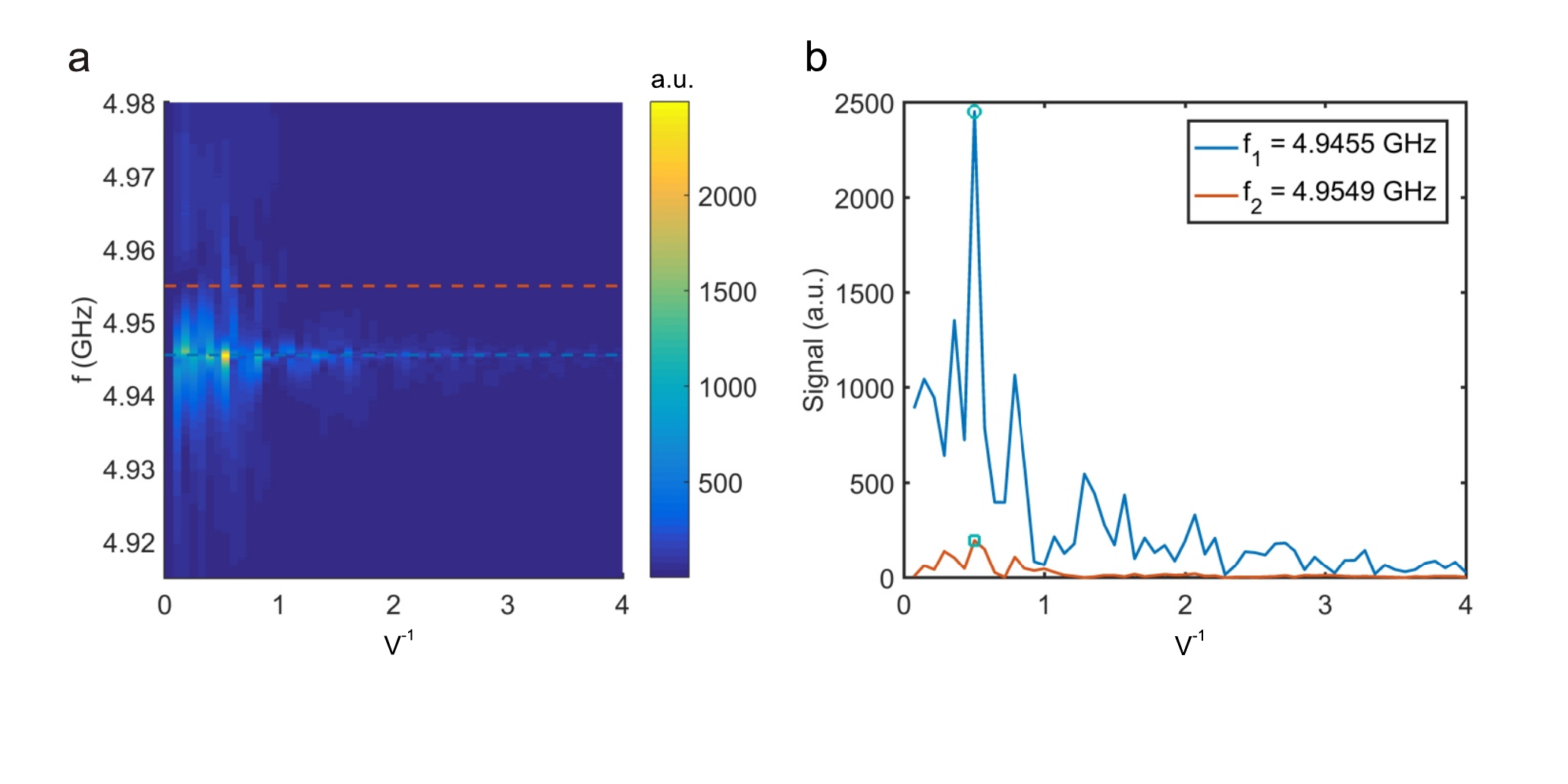}
\caption{(a) Fourier spectrum of data shown in Fig. \ref{Fig4}(a). (b) Linecuts at two different frequencies in (a), indicated by the dashed lines in (a) with corresponding colors. Note that the red trace is the FT of the data in Fig. \ref{Fig4}(c). The green circle in each linecut indicates the maximal FT signal.}
\label{Fig5}
\end{figure*}

We further investigate the flux-tunability of the SQUID in more details as shown in Fig. \ref{Fig4}(a), where cavity's resonant frequency is measured as a function of bias voltage applied to the Z-control line at P = -20 dB. The cavity frequency shows a periodic oscillation with bias voltage, resembling the behavior of conventional transmon made of Al/Al$_2$O$_3$-based SQUID \cite{Chow2010}. Note that there is a vertical stripe mode at V$_b$ $\approx$ -1.8 V, which is of unknown origin and not periodic in flux thus not related to the SQUID. It could be due to an accidentally formed magnetic impurity or field-tunable two-level system with energy coincides with cavity frequency at that bias voltage. Fig. \ref{Fig4}(b) show the linecuts along the frequency axis in Fig. \ref{Fig4}(a), with V$_b$ varied between V$_b$ = 2.5 V and 4.3 V in an equal step of 0.2 V [range marked by red triangles in Fig. \ref{Fig4}(a)]. In this bias range, the peak position moves away and restores, which corresponds one period of flux tuning and allows us to obtain the maximal tunability $\Delta\chi/2\pi$ $\approx$ 1.5 MHz. Using $\chi = g^2/2\pi(f_r - f_t)$, with the lowest $\chi/2\pi$ $\approx$ 8.5 MHz and the highest $f_r$ $\approx$ 4.9465 GHz, we can extract the maximum transmon frequency $f_{t, max}$ $\approx$ 6.53 GHz. The transmon frequency based on a symmetric SQUID oscillates against flux and can be approximated as $f_t$ $\approx$ $f_{t,max}\sqrt{cos(\pi\Phi/\Phi_0)}$, where $\Phi_0$ = $h/2e$ is the flux quantum and $\Phi$ is the magnetic flux threading the loop of SQUID \cite{Krantz2020}. Thus, the transmon frequency $f_t$ can be inferred as shown in green solid lines in Fig. \ref{Fig4}(a). In Fig. \ref{Fig4}(c), we show the linecut along the bias direction at a frequency of 4.9549 GHz [see the blue triangle in Fig. \ref{Fig4}(a)], in order to obtain more insights on the flux tuning. In a topological material with helical electronic states, the supercurrent flowing in its Josephson junction can be a combination of both 4$\pi$- and 2$\pi$-periodic phase term \cite{Li2018a,Wang2018a,Badiane2013}, namely $I_{J1(2)}$ = $I_{S1(2)}$[$A_{1(2)}sin\frac{\phi_{1(2)}}{2}$+$B_{1(2)}sin\phi_{1(2)}$], where the index 1(2) denotes junction 1 (2) in the SQUID, $sin\frac{\phi}{2}$ and $sin\phi$ represent the 4$\pi$ and 2$\pi$ components of the supercurrent, and $A$ and $B$ are the weight of each component in a junction. Note that $\phi_{1(2)}$ is the phase across junction 1(2) in the SQUID and has the relation of $\phi_{2}$ = $\phi_{1}$ + 2$\pi\frac{\Phi}{\Phi_0}$. Since the supercurrent through SQUID is a sum of $I_{J1}$ and $I_{J2}$, it has the form of $I_{SQUID}$ = $I_{S1A}sin\frac{\phi_{1}}{2}$ + $I_{S1B}sin\phi_{1}$ + $I_{S2A}sin(\frac{\phi_{1}}{2}+\frac{\pi\Phi}{\Phi_0})$ + $I_{S2B}sin(\phi_1+\frac{2\pi\Phi}{\Phi_0})$. Here, $I_{S1A(B)}$ denote the the weight of 4$\pi$ (2$\pi$) supercurrent component in junction 1, while $I_{S2A(B)}$ denote the the weight of 4$\pi$ (2$\pi$) supercurrent component in junction 2, respectively. Because the weight of 4$\pi$ and 2$\pi$ component in each junction is unknown, $I_{S1A}$, $I_{S1B}$, $I_{S2A}$ and $I_{S2B}$ are undetermined parameters. Considering the terms associated with $\Phi$, the overall supercurrent is single-period in flux tuning if one of the components (i.e., either $I_{S2A}$ or $I_{S2B}$) dominates the supercurrent. However, if the weight of 2$\pi$ and 4$\pi$ component is similar (i.e., $I_{S2A}$ $\approx$ $I_{S2B}$), the supercurrent can exist a flux oscillation with unequal spacing between neighboring peaks. This single-period and non-single-period behavior in supercurrent should also reflect on the flux modulation of cavity frequency. In order to gain more insight on the periodicity, we have performed Fourier transform (FT) on the 2D diagram of Fig. \ref{Fig4}(a) and the trace shown in Fig. \ref{Fig4}(c), and the results are shown in Fig. \ref{Fig5}(a) and Fig. \ref{Fig5}(b), respectively. Note we also performed FT on the data shown in Fig. \ref{Fig3}(c) for comparison, as presented in section VI in the supplementary material. The 2D map of FT in Fig. \ref{Fig5}(a) and the linecuts at both f = 4.9455 GHz and f = 4.9549 GHz in Fig. \ref{Fig5}(b) indeed show a dominant spectral component at around 0.5 V$^{-1}$, which corresponds to the oscillation period $\Delta$V $\approx$ 2V in Fig. \ref{Fig4}(c). The FT signals in Fig. \ref{Fig5}(a) show some level of broadness around 0.5 V$^{-1}$, possibly due to the glitch noises in Fig. \ref{Fig4}(a) [i.e., around bias voltage -4.5 V and 5.5 V]. In a topological junction, the 2$\pi$-periodic supercurrent (bulk transport) is expected to inevitably contribute to the overall supercurrent, while the 4$\pi$ supercurrent often requires careful tuning to appear \cite{Wang2018a,Li2018a}. It is also worth noting that the 4$\pi$-periodic supercurrent tends to restore 2$\pi$ periodicity, as a result of various relaxation processes \cite{Kwon2004,San-Jose2012,Rainis2012}. These relaxation mechanisms should play a role in DC measurements, and are in fact the reason that RF SQUIDs were adopted to probe the existence of 4$\pi$-periodic supercurrent in topological junctions \cite{Bocquillon2017,Li2018a}. Although our measurement is based on microwave techniques operating at GHz, the integrated measurement time is around seconds, which may still suffer from the relaxation processes happening at a time scale around microseconds \cite{Badiane2013,Rainis2012}. Thus, we attribute the component at around 0.5 V$^{-1}$ to the 2$\pi$-periodic supercurrent. On the other hand, we do not see an evident FT signal at around 0.25 V$^{-1}$ which corresponds to 4$\pi$-periodic supercurrent (see also section VI for similar analysis on the FT of Fig. \ref{Fig3}(c)). In future endeavor, a device with gates to tune between trivial bulk state and topological surface state \cite{Wang2018a}, which allows one to switch supercurrent between 2$\pi$- and 4$\pi$-period, would help to understand whether these relaxation processes play a role in this type of measurement.

\maketitle
\section{IV. Summary}

In summary, we have fabricated and characterized a transmon-like superconducting quantum circuit made of Weyl semimetal MoTe$_{2}$ and superconducting cavity of MoRe. The successfully made MoTe$_{2}$ Josephson junction revealed itself in the qubit power-dependent measurement. The power-dependent cavity frequency shift allows us to extract the transmon frequency, Josephson energy and the critical current flowing through the SQUID. The cavity frequency is tunable with magnetic flux applied to the loop of SQUID area, indicating the flux modulation of transmon frequency. The flux-tuning shows a single-period oscillation, suggesting the 2$\pi$-phase term is the dominating contribution in the supercurrent or the measurement is mediated by the relaxation processes which tend to restore the 4$\pi$ to 2$\pi$ periodicity in a topological junction. Our study demonstrates a flux-tunable SQUID quantum circuit based on topological material. The transmon-like geometry provide a platform for use in quantum information processing in the future.

\maketitle
\section{V. Acknowledgments}
K. L. Chiu would like to acknowledge Karl Petersson, Tongxing Yan, Peng Duan, Fei Yan and Siyuan Han for the useful discussions on data analysis. The authors would like to thank the funding support from Guangdong Provincial Key Laboratory (Grant No. 2019B121203002), National Natural Science Foundation of China (Grant No. 11934010) and Ministry of Science and Technology of Taiwan (Grant No. MOST 109-2112-M-110-005-MY3). 

\maketitle
\section{VI. Author contributions}
K.L.C conceived the project. D.G.Q fabricated the devices with contributions from K.L.C, Z.T.Z, S.L, Y.Z and D.P.Y. K.L.C performed the measurements with contributions from J.W.C and W.Y.L. K.L.C, V.M and D.T analyzed the data. K.L.C wrote the manuscript with input from all the co-authors.

\maketitle
\section{VII. Declaration}
The order of affiliation does not represent the contributing order, Shenzhen Institute for Quantum Science and Engineering and National Sun Yat-Sen University are the co-first institute in this work. Device fabrications and measurements were performed in Shenzhen Institute for Quantum Science and Engineering, while data process and analysis were carried out in National Sun Yat-Sen University.


 \widetext
 \clearpage
 \begin{center}
 	\textbf{\large Supplementary materials}
 \end{center}

 \setcounter{section}{0}
 \setcounter{equation}{0}
 \setcounter{figure}{0}
 \setcounter{table}{0}
 \setcounter{page}{1}
 \makeatletter
 \renewcommand{\thesection}{S\arabic{section}}
 \renewcommand{\thesection}{\Roman{section}}
 \renewcommand{\theequation}{S\arabic{equation}}
 \renewcommand{\thefigure}{S\arabic{figure}}
 \renewcommand{\bibnumfmt}[1]{[S#1]}

 \section{I. Device fabrication}

\begin{figure}[!t]	
 	\includegraphics[scale=0.69]{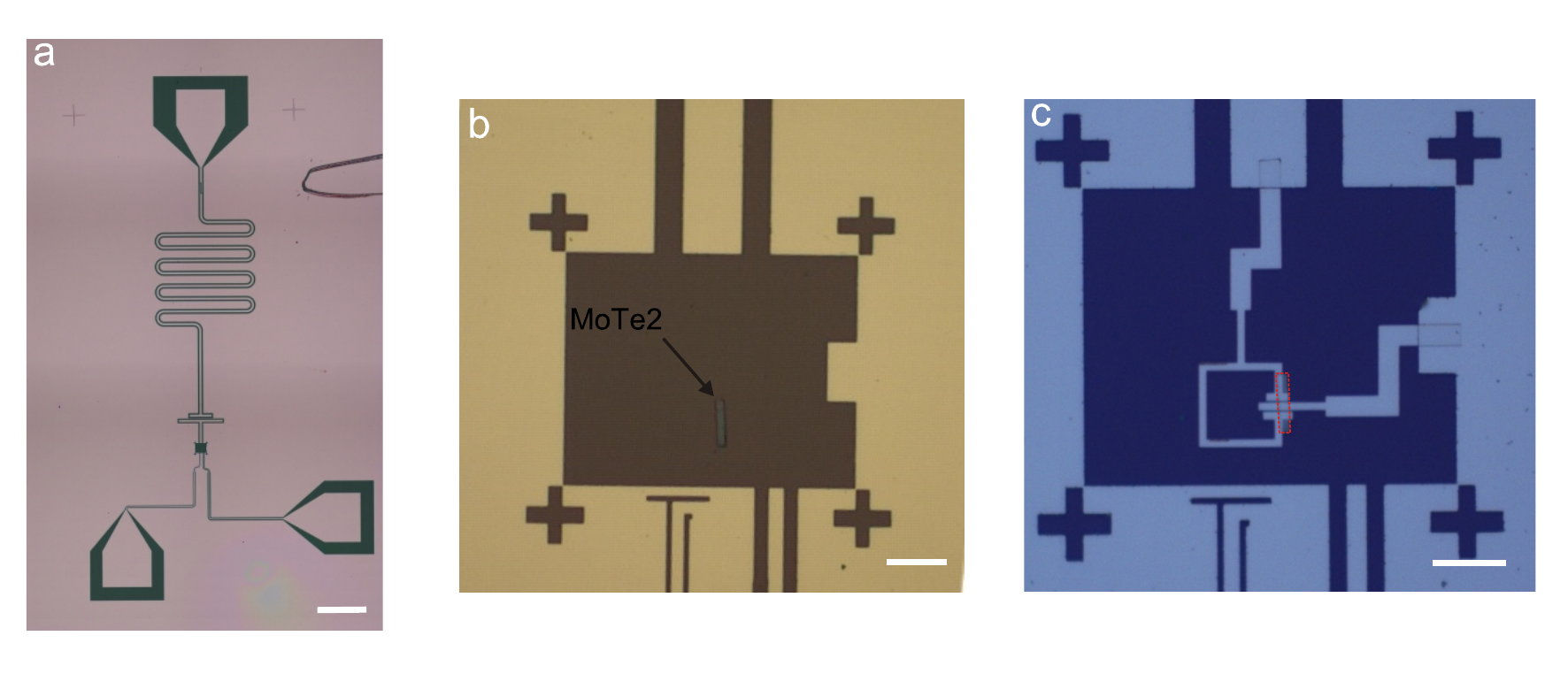}
 	\caption{Optical micrograph of the device (a) after the fabrication of superconducting cavity and feedlines. The scale bar is 450 $\mu$m. (b) after a strip of MoTe$_2$ was transferred onto the square area. The scale bar is 20 $\mu$m. (c) after the fabrication of the SQUID. The scale bar is 20 $\mu$m.}      
 	\label{FigS1}
 \end{figure}

The general steps for device fabrication has been specified in the main text. Here, we provide more details during different stages in fabrication. After the sputtering of 80 nm MoRe (Ar flow: 50 sccm, Pressure: 0.57 Pa, Rate: 3.3 A/s) onto the intrinsic Si wafers capped with 285 nm SiO$_2$, the structures of cavity and feedlines are defined by a laser writer (Resist: S1813, Developer: MIF 319). After the ICP etching (SF6: 40 sccm, Ar: 10 sccm), the optical micrograph of the as-made cavity and control lines is shown in Fig. \ref{FigS1}(a). Afterwords, a strip of multilayered MoTe$_2$ was transferred onto the square area, as shown in Fig. \ref{FigS1}(b), using a dry-polymer based process \cite{Wang2013}. Note that all the exfoliation and transfers were performed in a N$_2$ filled glove box to prevent the air-sensitive T$_d$-MoTe$_2$ from degradation. The device was immediately covered by PMMA after the completion of transfer. After a writing of electron-beam lithography and sputtering of superconducting leads (NbN), the SQUID is fabricated as shown in Fig. \ref{FigS1}(c). Note that we do not do the ion mill between the superconducting leads and cavity, because the oxidation of MoRe is negligible during our fabrication processes.

\section{II. Measurement scheme}

\begin{figure}[h]
 	\includegraphics[scale=1]{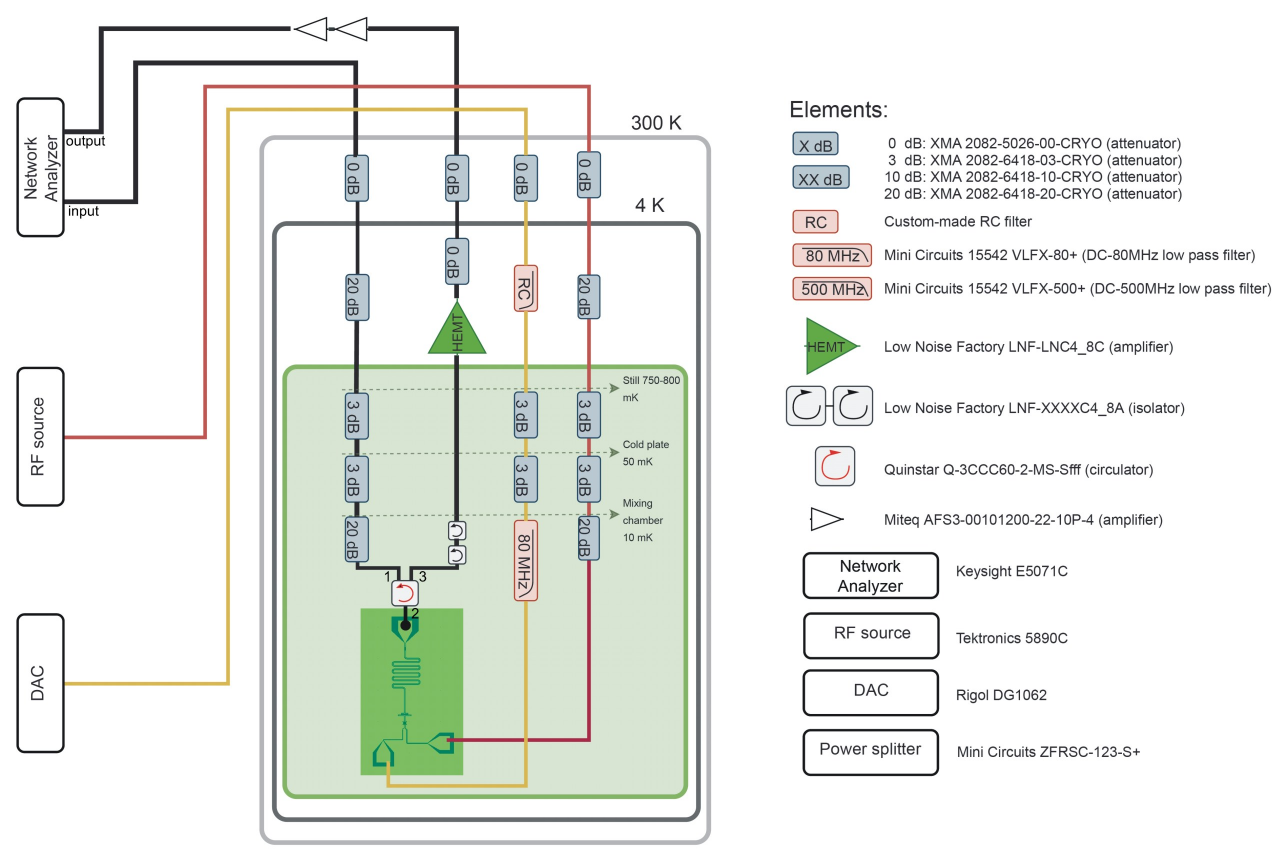}
 	\caption{The measurement scheme used to perform the power-dependence and flux-tuning of cavity's resonant frequency. The right panel shows the detail of the elements.}      
 	\label{FigS2}
 \end{figure}

Fig. \ref{FigS2} shows the measurement scheme used in this work. All the experiments detailed in the main text were performed in a dilution refrigerator with a base temperature of 10 mK. The devices were bonded in a light tight Al box which is thermally anchored to the mixing chamber and enclosed in a Al cylinder shield to prevent flux noises. The scheme showns in Fig. \ref{FigS2} is adopted to perform the power-dependence and flux-tuning of cavity's resonant frequency. In this setup, four coaxial lines were used to control the device: two are assigned for cavity readout, one for flux control and one for drive tone. The cavity readout line represented by black solid line in Fig. \ref{FigS2} consists of an input and an output line, as shorted by readout-in and readout-out in the following content. The readout-in line connected to the input of a network analyzer (Keysight E5071C) is heavily attenuated in low temperature to reduce noise and thermal excitation before connecting to a circulator (port 1). The second port of the circulator is connected to the reflective cavity of device. The reflected readout signal is out through the third port of circulator and goes into the readout-out line. The readout-out line was connected to two isolators in series to shield the device from thermal radiation from the HEMT amplifier on the 4K stage. The HEMT (LNF-LNC4 8C) has about 40 dB of gain and a noise temperature of 2 K. The readout signal is further amplified by two room-temperature amplifiers before going into the output of the network analyzer. The flux-tuning line represented by yellow solid line in Fig. \ref{FigS2} was connected to a DAC (Rigol DG 1062) to apply DC voltage. This line is heavily filtered by a home-made RC filter at 4 K and a 80 MHz low-passed filter at 10 mK to reduce the thermal noises of frequencies greater than 300 kHz. The applied DC voltage to the 1.5 k$\Omega$ resistor in the RC filter allows us to generate a DC current in device's Z-control line for flux tuning. The drive-tone line was connected to a RF source (Tektronic 5890c) in room temperature and heavily attenuated in low temperature, again to reduce the unwanted noises, as shown in the red solid line in Fig. \ref{FigS2}.

\section{III. Power dependence}
 
\begin{figure}[!t]	
 	\includegraphics[scale=0.7]{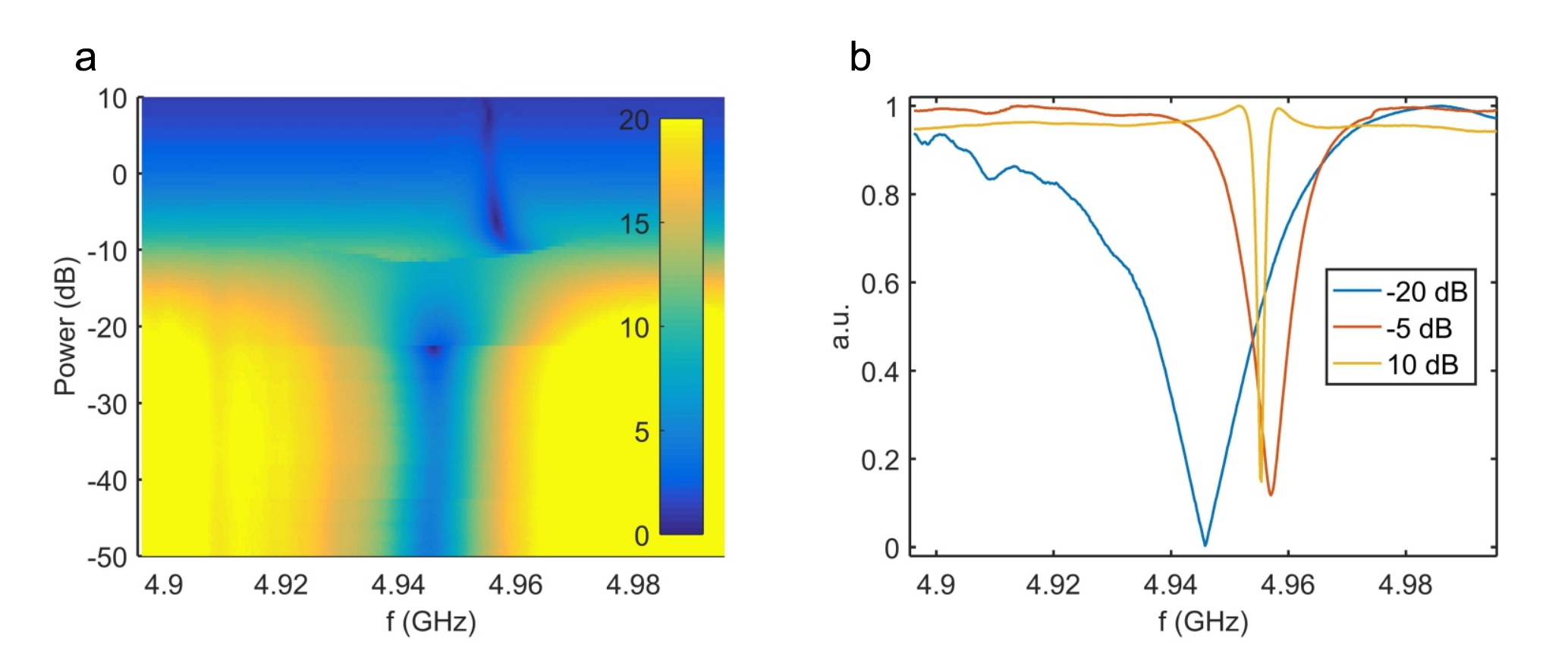}
 	\caption{The power spectrum measured by reducing the attenuation of 3 dB in the readout-in line as compare to Fig. \ref{Fig2}. (a) The power spectrum measured by recording $S_{21}$ as a function of power and frequency applied to the readout-in line. (b) Linecuts in (a) at three different powers (data normalized to 1).}      
 	\label{FigS3}
 \end{figure}

For a conventional transmon qubit, Bishop $et$ $al.$ developed a semiclassical model under the bad cavity and strong-dispersive approximations \cite{Bishop2010}: 

\begin{eqnarray}
\chi (A) &=& \sigma_z \frac{g^2}{\sqrt{2g^2 (A^2 + \sigma_z) + \delta^2}}, \label{eqn S1}
\end{eqnarray}
where $\chi$ is the cavity dispersive shift, $A$ is cavity amplitude (drive amplitude), $\delta$ is the cavity-qubit detuning, $g$ is the coupling strength between the cavity and qubit, and $\sigma_z$ is Pauli matrix. This equation satisfies the relations $\chi (0) \approx \pm g^2 /\delta$ and $lim_{A\rightarrow\infty}\chi (A)$ = 0, meaning the anharmonicity of the cavity (e.g. the difference in $\chi$ for $N$ and $N + 1$ excitations) is maximal at low power and continuously diminish as the drive strength is increased. In a conventional transmon qubit \cite{Reed2010}, the cavity response is linear at the zero-power limit, but as the drive strength increases, the turn-on of anharmonicity would distort the line shape and result in dips, which Bishop attributes to the ensemble-averaging of the bistalbe solutions of amplitude $A$ in Eq. \ref{eqn S1} \cite{Bishop2010}. In our case, we attribute the broad regime, i.e. -15 dB $\leq$ P $\leq$ -10 dB in Fig. \ref{FigS3} (a), to the destructive interference of the two semiclassical solutions of $A$ with almost opposite phases. Finally, with the drive strength further increases, the frequency extent of the bistable region shrinks and eventually cavity shift into resonance with the drive. At this point, the cavity population rapidly grow while its anharmonicity shrinks – thus reaching the bright state.

\section{IV. coupling strength estimation}
\begin{figure}[!t]	
 	\includegraphics[scale=0.45]{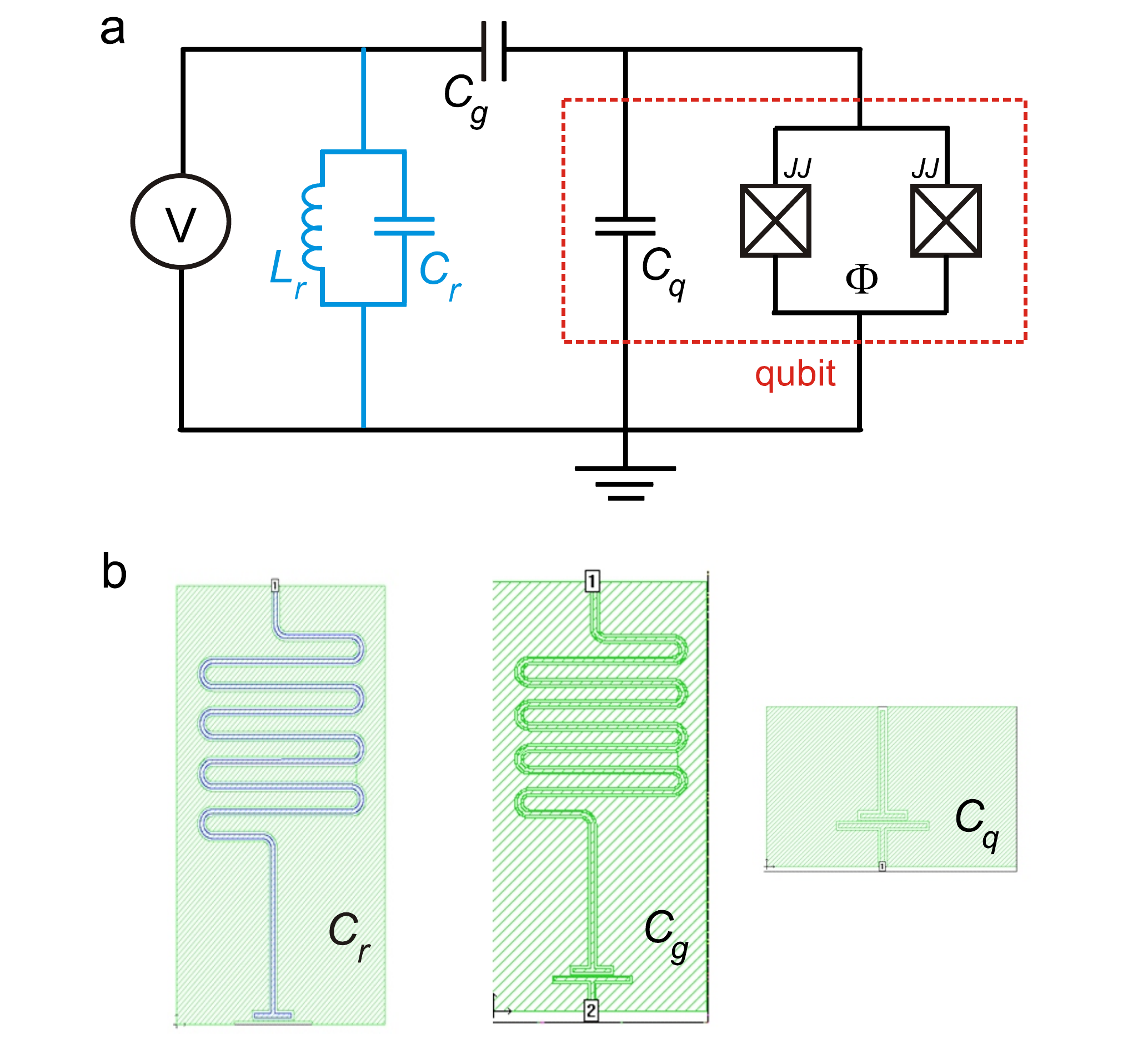}
 	\caption{(a) Effective circuit diagram of our transmon-like SQUID device. The red dashed square highlights the SQUID and shunted capacitor, which are effectively denoted as qubit in the diagram. The blue solid lines denote the readout cavity, where $C_r$ ($L_r$) is the effective capacitance (inductance) of the cavity. $C_g$ is the capacitance between the readout cavity and T-shape island while $C_q$ is that between the T-shape island and ground. (b) The geometries used in SONNET simulation, from which we obtained $C_q$ = 82 fF, $C_g$ = 5 fF and $C_r$ = 1.4 pF.}      
 \label{FigS4}
 \end{figure}

Fig. \ref{FigS4} shows the effective circuit diagram of our transmon-like SQUID device. The two Josephson junctions are shunted by the capacitance $C_q$ between the T-shape island and ground. $C_g$ is the capacitance between the readout cavity and T-shape island while $C_r = \epsilon_r\pi/2\omega_rZ_0$ ($L_r$) is the effective capacitance (inductance) of the readout cavity, where $\omega_r = 1/\sqrt{L_r C_r}$ is the cavity resonant frequency. Substituting $Z_0$ = 50, $\epsilon_r$ = 1 and $\omega_r$ = 2$\pi$ $\times$ 4.955 GHz, we get $C_r$ = 1 pF, which is close to the simulated value 1.4 pF [leftmost panel in Fig. \ref{FigS4} (b)]. For other parameters, we use Sonnet Software to simulate the capacitance to get $C_g$ = 5 fF and $C_q$ = 82 fF, as shown in Fig. \ref{FigS4} (b). The qubit-cavity coupling strength $g$ can be expressed as $g$ = $\beta\sqrt{\frac{2e^2\omega_r}{\hbar C_r}}$, where $\beta$ = $\frac{C_g}{C_q + C_g}$ \cite{Koch2007}. Substituting all the parameters we have $g/2\pi$ = 116 MHz, which is in agreement with the value in a similar design in Ref. \cite{Larsen2015}.

\section{V. Two-tone measurements}

\begin{figure}[!t]	
 	\includegraphics[scale=0.75]{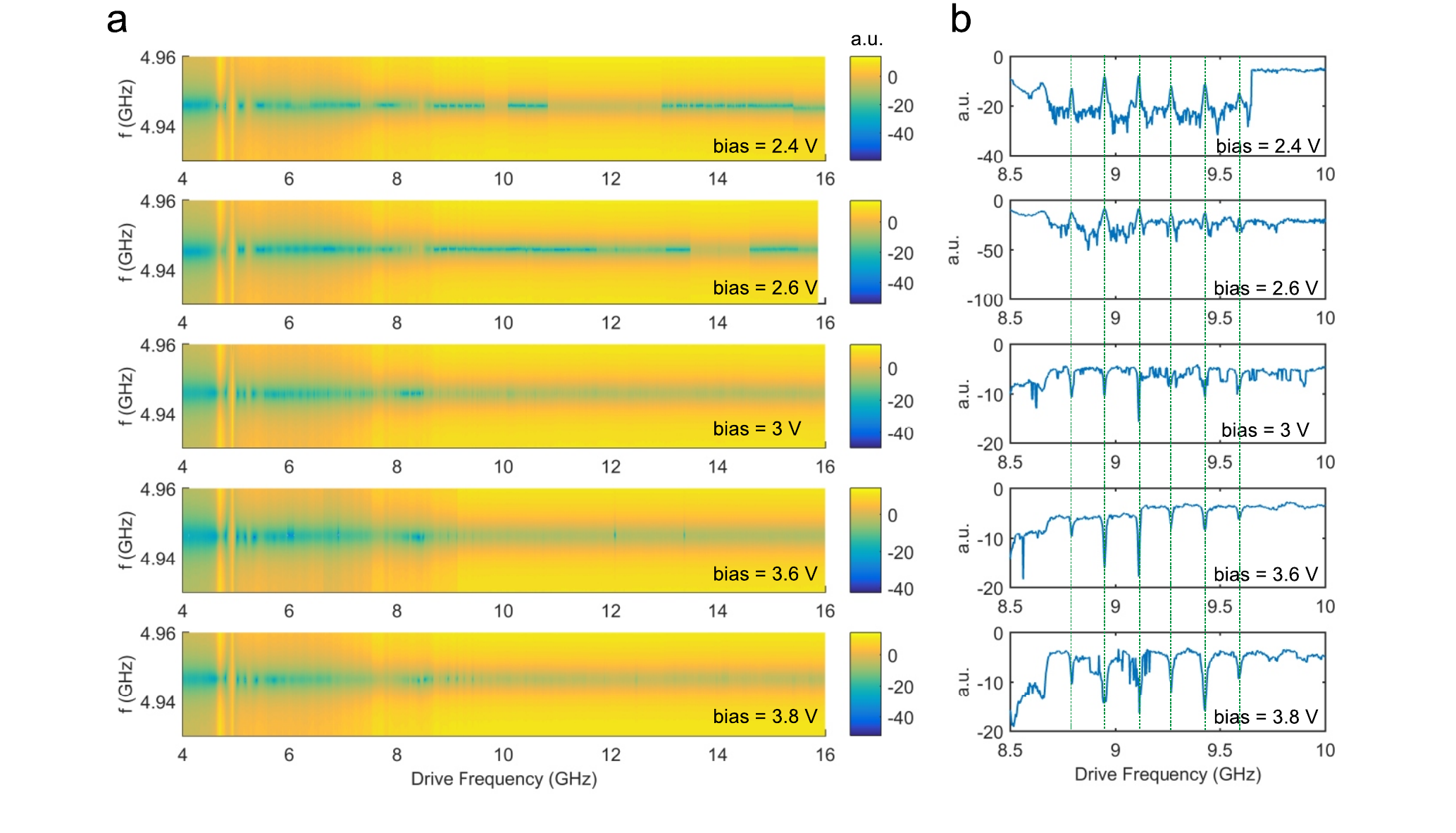}
 	\caption{Two-tone spectroscopy. (a) S$_{21}$ measured as a function of readout-in frequency (y-axis) and drive frequency (x-axis) at different bias voltages ranging from 2.4 V to 3.8 V. The readout-in tone power is -22 dB and the drive tone power is -10 dB. (b) Line cuts for each bias voltage at cavity resonant frequency and in a range of drive frequency from 8.5 GHz to 10 GHz. The green dashed lines indicate the peaks possibly associated with standing wave in the coaxial cables.}      
 	\label{FigS5}
 \end{figure}

\begin{figure}[!t]	
 	\includegraphics[scale=0.65]{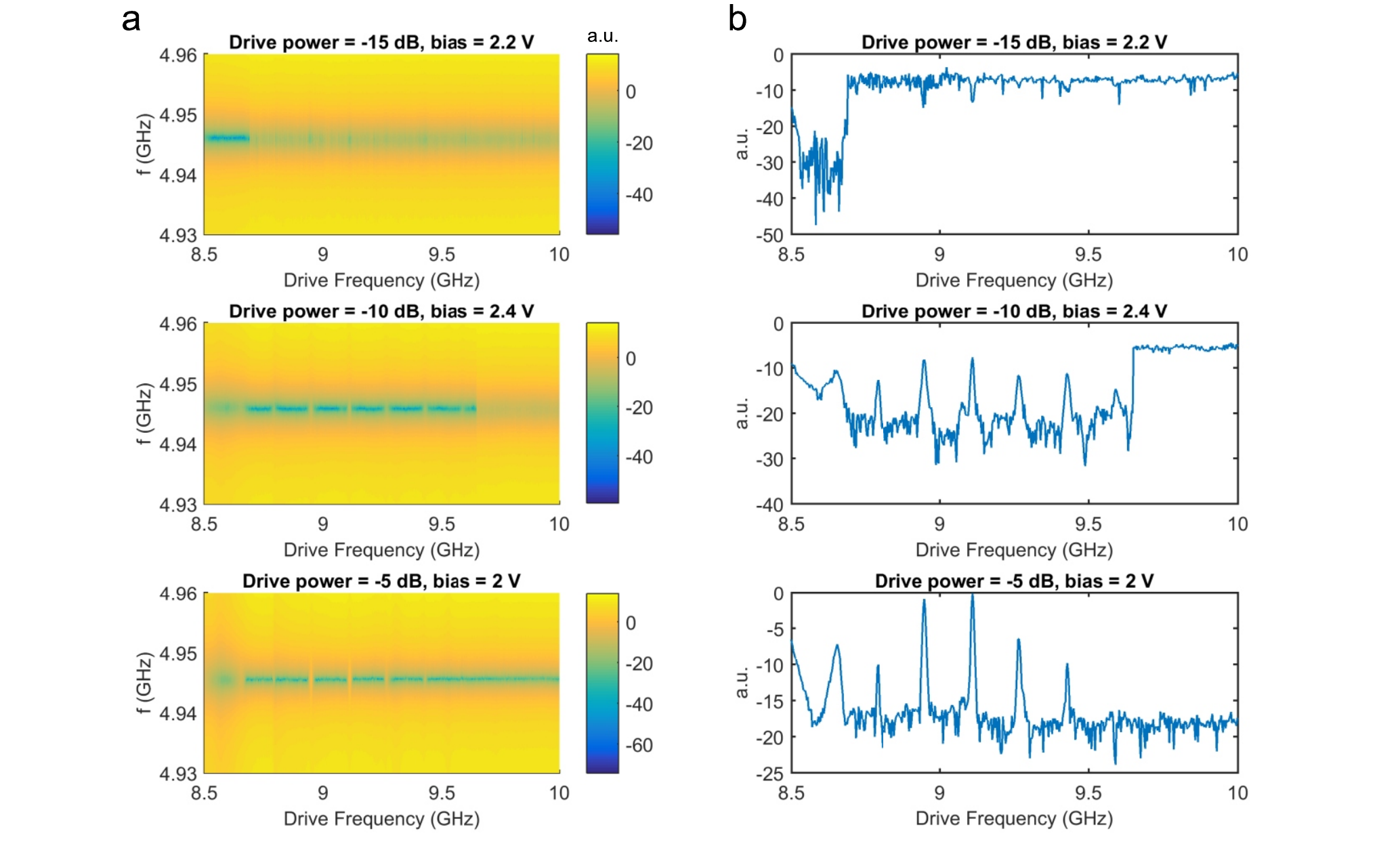}
 	\caption{Power-dependent periodic oscillations. (a) S21 measured as a function of readout-in frequency (y-axis) and drive frequency (x-axis) at different drive tone powers (-15 dB, -10 dB and -5 dB) and bias voltages. The readout-in tone power is -22 dB. (b) Line cuts of (a) at cavity resonant frequency, showing the periodic oscillations are power-dependent.}      
 	\label{FigS6}
 \end{figure}

Two-tone spectroscopy was performed by measuring S$_{21}$ as a function of frequency of a tone applied to the readout cavity and another tone applied to the XY-control line. Our result is shown in Fig. \ref{FigS5}(a), in which the y-axis is the readout frequency measured by network analyzer while the x-axis is the drive frequency generated by a RF source. We have performed this measurement at different applied biases ranging from 2.4 V to 3.8 V. However, no shift in cavity frequency was observed in the whole probed range of drive frequency, indicating a very short coherence time. As seen in Fig. \ref{Fig4}(a), the fact that the transmon frequency crosses the cavity periodically but does not form avoided crossings, indicates the device is not in the strong coupling regime (i.e., $\Omega$ $\gg$ $\gamma$, $\kappa$, where $\Omega$ is the transmon-cavity energy exchange rate and $\gamma$($\kappa$) is transmon (cavity) decay rate). In this case, our transmon should have a decay rate $\gamma$ $>$ $\Omega$ = $g/\pi$ = 232 MHz, hence a coherence time 1/$\gamma$ $<$ 4.3 ns. In the experiment of energy level spectroscopy, the linewidth of the spectroscopic peak is $w$ = $\frac{1}{\pi T^{\ast}_2}$ \cite{Schmitt2020,Schuster2005}. Thus, for a device with coherence less than 4 ns, the linewidth would be at least greater than 80 MHz. One would expect to see only a very hazy spectroscopy line, since the integrated area of the resonance line is a fixed quantity, and if it is spread over more than 80 MHz, it's value at the center frequency would be largely indistinguishable from the majority of the linewidth. In our measurements, we observed a periodic feature in the lincuts at cavity resonant frequency and in the range between 8.5 GHz and 10 GHz in drive frequency [Fig. \ref{FigS5}(b)]. These oscillating peaks appear at all bias voltages and their position is independent of the applied biases (i.e., transmon frequencies), as indicated by the green dashed lines in Fig. \ref{FigS5}(b). In addition, the resonances were only observable at very high power, as shown in Fig. \ref{FigS6}. These oscillations may be associated with a standing wave ($\Delta$f $\approx$ 159 MHz, $\lambda$ $\approx$ 1.25 m) in the coaxial cables, as also being reported in Ref. \cite{Kroll2018a}.

\section{VI. Flux periodicity analysis}
\begin{figure}[!t]	
 	\includegraphics[scale=0.73]{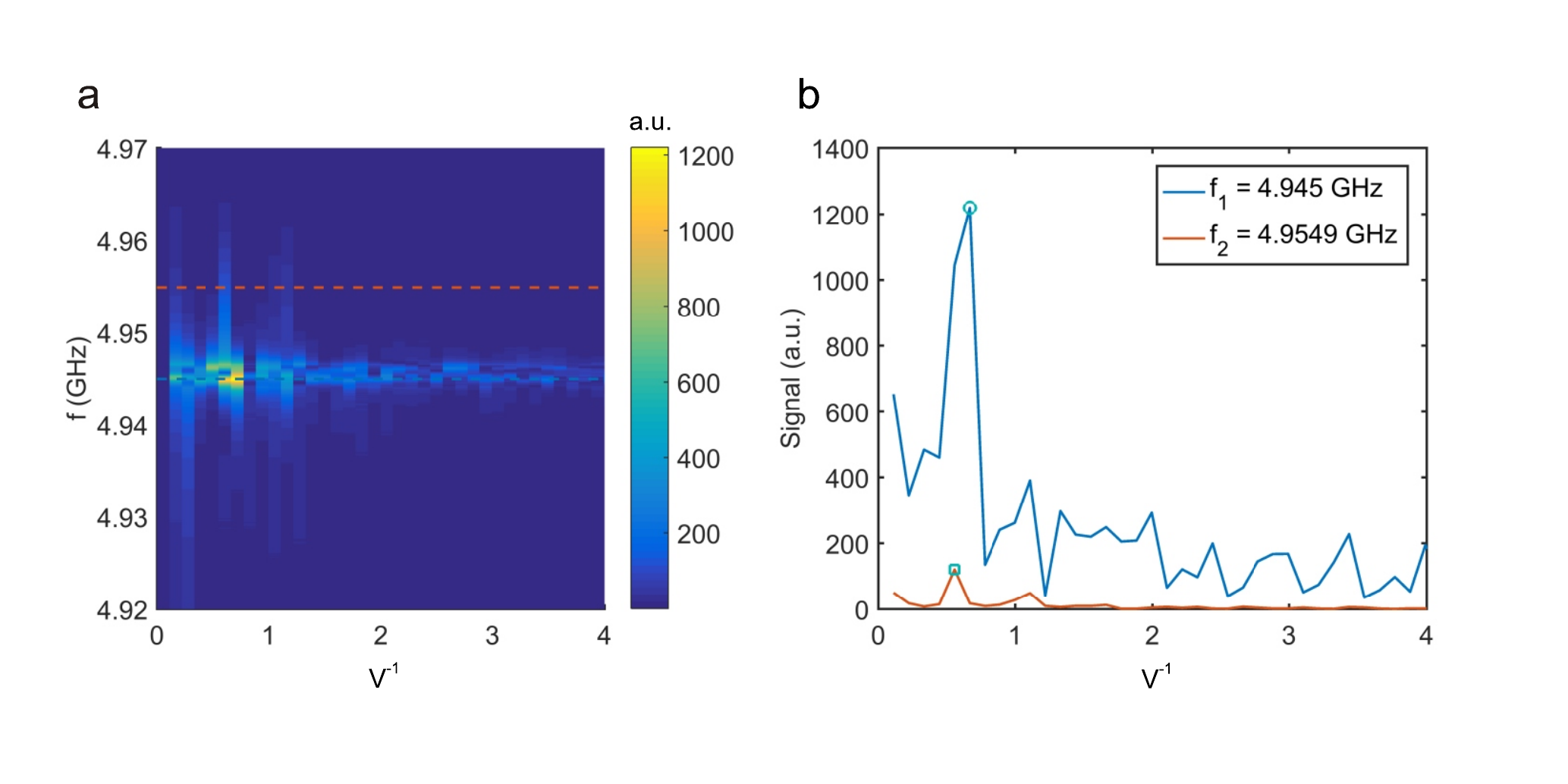}
 	\caption{(a) Fourier spectrum of data shown in Fig. \ref{Fig3}(c). (b) Linecuts at two different frequencies in (a), indicated by the dashed lines in (a) with corresponding colors. The green circle indicates the maximal FT signal in each linecut.}      
 	\label{FigS7}
 \end{figure}

The Fourier transform (FT) of Fig. \ref{Fig3}(c) is shown in Fig. \ref{FigS7}. The dominant spectral component at around 0.5 V$^{-1}$ is also observed, which is consistent with the observations in Fig. \ref{Fig5}. However, there appears to be a non-ignorable component around 1.1 V$^{-1}$, which survives even at higher frequencies. In our topological junctions, we anticipate the 2$\pi$-periodic supercurrent is an inevitable contribution and should correspond to the 0.5 V-1 in both Fig. \ref{Fig5} and Fig. \ref{FigS7}. As depicted in the main text, the 2$\pi$-periodic supercurrent has the flux relation of $I_{S2B}$$sin(\phi_1+\frac{2\pi\Phi}{\Phi_0})$, while the 4$\pi$-periodic supercurrent has that of $I_{S2A}$$sin(\frac{\phi_{1}}{2}+\frac{\pi\Phi}{\Phi_0})$. If the FT component at 0.5 V$^{-1}$ corresponds to the 2$\pi$ supercurrent, the 4$\pi$ supercurrent should appear at around 0.25 V$^{-1}$, not 1.1 V$^{-1}$. Therefore, we conclude that the signal at 1.1 V$^{-1}$ in Fig. \ref{FigS7} does not represent the 4$\pi$-periodic supercurrent. On the contrary, if the FT signal at 0.5 V$^{-1}$ in both Fig. \ref{Fig5} and Fig. \ref{FigS7} corresponds to the 4$\pi$ supercurrent, there should be a signal at around 1 V$^{-1}$ corresponding to 2$\pi$ supercurrent, but it is not observed in Fig. \ref{Fig5}.

\end{document}